\documentclass[prb,aps,epsf,twocolumn,showpacs]{revtex4-1}
\usepackage[pdftex]{graphicx}
\usepackage{dcolumn}
\usepackage{bm}
\usepackage{epsfig}
\usepackage{latexsym}
\usepackage{amsmath}
\usepackage{color}
\usepackage{array}

\newcommand{\e}{\varepsilon}

\renewcommand{\>}{\rangle}
\renewcommand{\(}{\left(}
\renewcommand{\)}{\right)}
\renewcommand{\[}{\left[}
\renewcommand{\]}{\right]}
\renewcommand{\v}[1]{\mathbf{#1}} 

\begin{document}
\pdfoutput=1
\title{Topological Superconductivity and Majorana Fermions in Metallic Surface-States}
\author{Andrew C. Potter}\author{Patrick A. Lee}
\affiliation{Massachusetts Institute of Technology 77 Massachusetts Ave. Cambridge, MA 02139}
\begin{abstract} 
Heavy metals, such as Au, Ag, and Pb, often have sharp surface states that are split by strong Rashba spin-orbit coupling.  The strong spin-orbit coupling and two-dimensional nature of these surface states make them ideal platforms for realizing topological superconductivity and Majorana fermions.  In this paper, we further develop a proposal to realize Majorana fermions at the ends of quasi-one-dimensional metallic wires.  We show how superconductivity can be induced on the metallic surface states by a combination of proximity effect, disorder, and interactions.  Applying a magnetic field along the wire can drive the wire into a topologically non-trivial state with Majorana end-states.  Unlike the case of a perpendicular field, where the chemical potential must be fined tuned near the Rashba-band crossing, the parallel field allows one to realize Majoranas for arbitrarily large chemical potential.  We then show that, despite the presence of a large carrier density from the bulk metal, it is still possible to effectively control the chemical potential of the surface states by gating.  The simplest version of our proposal, which involves
only an Au(111) film deposited on a conventional superconductor, should be
readily realizable.
\end{abstract}
\maketitle


The observation that ordinary s-wave superconductivity (SC) coupled by proximity to a topological insulator can create an exotic topological superconductor possessing non-Abelian Majorana states\cite{FuKane} has inspired a large body of theoretical work, and spurred a new experimental thrust to realize non-Abelian particles in the laboratory.  While there are a variety of theoretical proposals\cite{FuKane,Fujimoto,SauSemiconductorSpinOrbit,AliceaPRB,LutchynWires,OregWires,PALee,ChungHalfMetal}, a particularly promising class involve combining conventional superonductors with two-dimensional materials with Rashba-type spin-orbit coupling (SOC)\cite{Fujimoto,SauSemiconductorSpinOrbit,AliceaPRB,LutchynWires,OregWires}.  

In these schemes, superconductivity (SC) is induced in a spin-orbit coupled material by proximity to an ordinary s-wave superconductor.  The two Rashba-split bands effectively convert the s-wave pairing into helical $p+ip$ and $p-ip$ pairing respectively.  Applying a magnetic field perpendicular to the plane opens up a Zeeman gap in the Rashba spectrum.  If the chemical potential is tuned within this Zeeman gap, then only one of the helically paired bands remains, leaving a single-helicity, $p+ip$ superconductor.  Superconductors with $p+ip$ pairing are known to be host non-Abelian Majorana bound-states in vortex cores\cite{Ivanov}, and at the ends of one-dimensional domains\cite{KitaevWire}.  Majorana end-states of one-dimensional wires are particularly interesting, since they can be manipulated by simple electrostatic gating\cite{AliceaNetworks,LutchynWires,ACPMultiBand1}.  Furthermore, networks of topological superconducting wires provide an effecient platform for braiding Majorana fermions to probe their non-Abelian statistics\cite{AliceaNetworks}.

So far, the bulk of theoretical and experimental efforts along these lines has focused on using semiconducting nanowires with heavy elements, such as InSb. Such semiconducting nanowires can be grown with very few defects and impurities.  These materials can also have very large electron g-factors (as high as $\approx 70$)\cite{StanescuDisorder}, allowing one to create a large Zeeman splitting in the nanowire with an external magnetic field, without significantly impacting the nearby superconductor\cite{LutchynWires,OregWires}.  However, semiconducting materials typically produce only very tiny Rashba SOC, $\Delta_{\text{so}}$.  For example, the Rashba SOC in InSb nano--wires is $\Delta_{\text{so}}\approx 1$K\cite{AliceaPRB,LutchynWires}.  Despite the presence of relatively heavy elements In and Sb, which have large atomic SOC, the Rashba component of SOC is very small, because it comes only from the breaking of inversion symmetry by the wire-superconductor interface.  Since these wires are typically quite large (10's-100's of nm in diameter), the electron wave-functions are spread over a large distance and do not strongly feel the inversion asymmetry from the interface.

Small spin-orbit coupling is problematic for two main reasons.  First the size of the p-wave superconducting gap protecting Majorana end-states is limited by $\Delta_{\text{p-wave}}\lesssim \Delta_{\text{so}}$\cite{ACPDisorder}, requiring one to work at very low-temperatures.  Second, small SOC renders the induced superconductivity extremely vulnerable to even very small amounts of disorder\cite{ACPDisorder,StanescuDisorder,BrouwerDisorder}.  The extreme sensitivity to disorder for small SOC may be problematic even though the bulk of semiconducting wires are typically quite clean; furthermore the superconducting gap will also be suppressed by roughness or inhomogeneity in the wire--superconductor interface.

\begin{figure*}[ttt]
\begin{center}
{\bf a)}\includegraphics[width=1.7in]{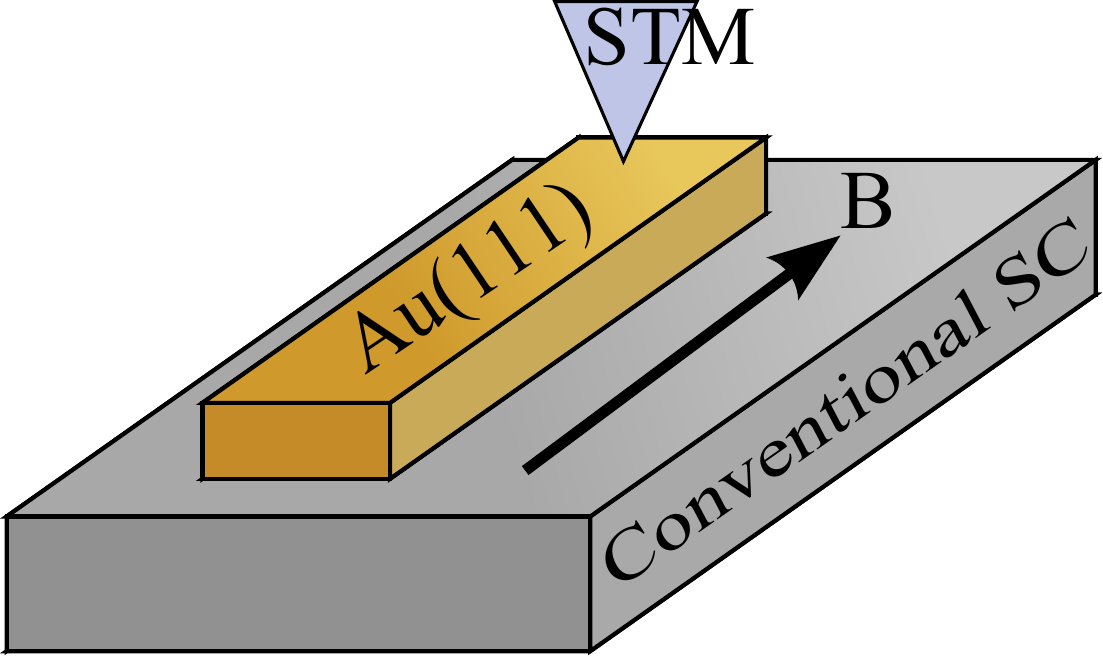}\hspace{-0.1in}
{\bf b)}\includegraphics[width=1.2in]{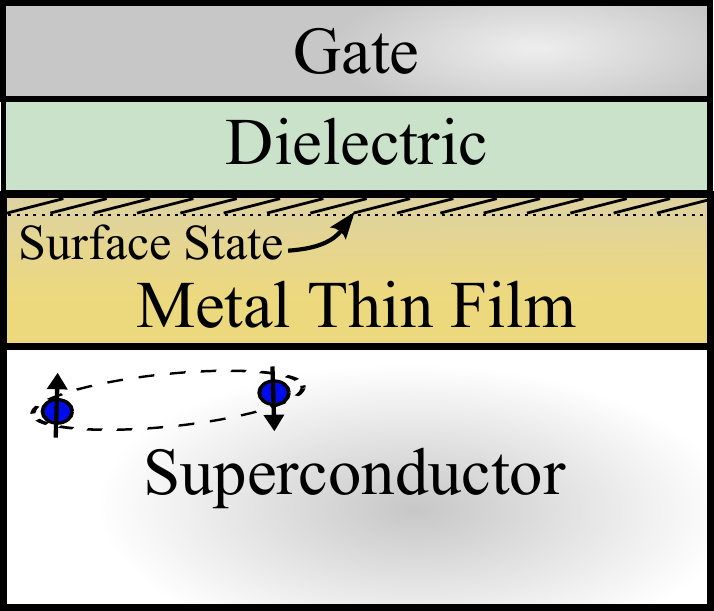}\hspace{0in}
{\bf c)}\hspace{-0.1in}\includegraphics[width=1.9in]{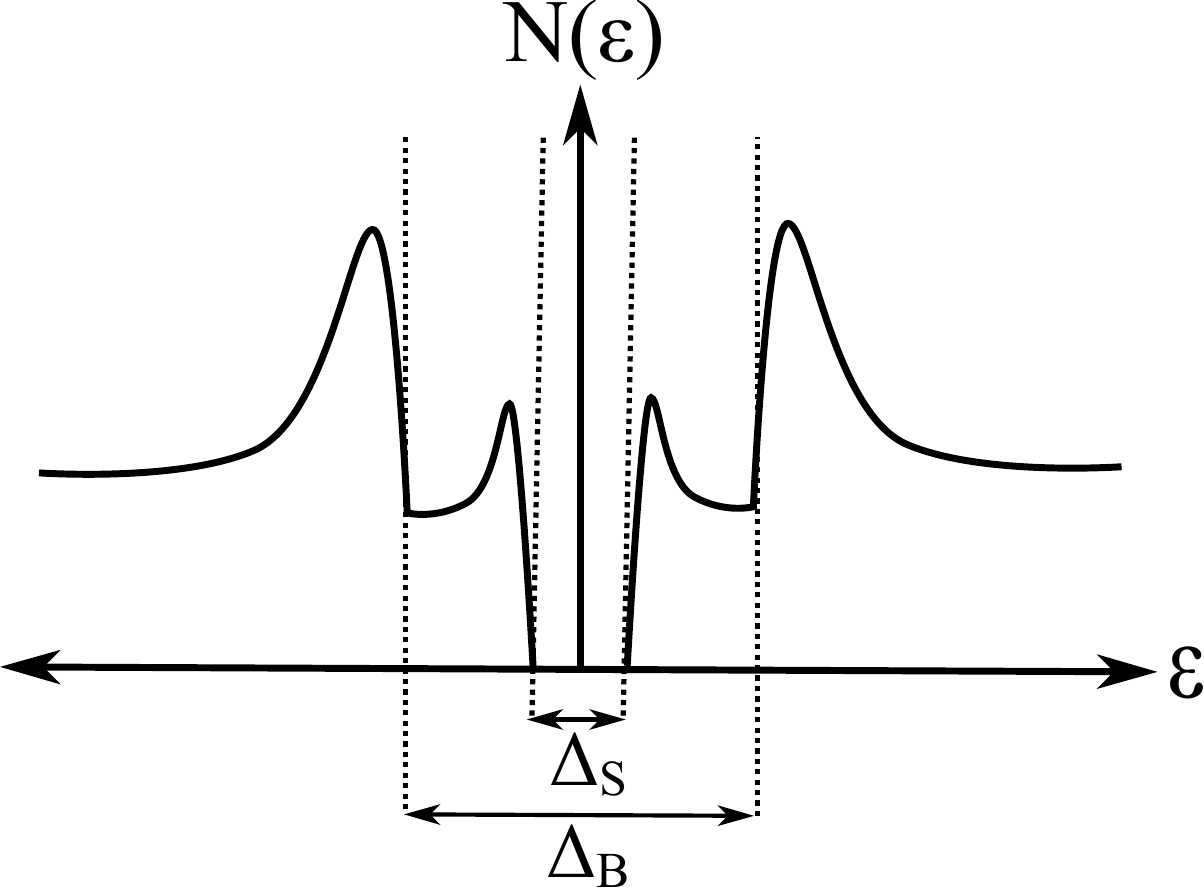}\hspace{-0.1in}
{\bf d)}\includegraphics[width=1.9in]{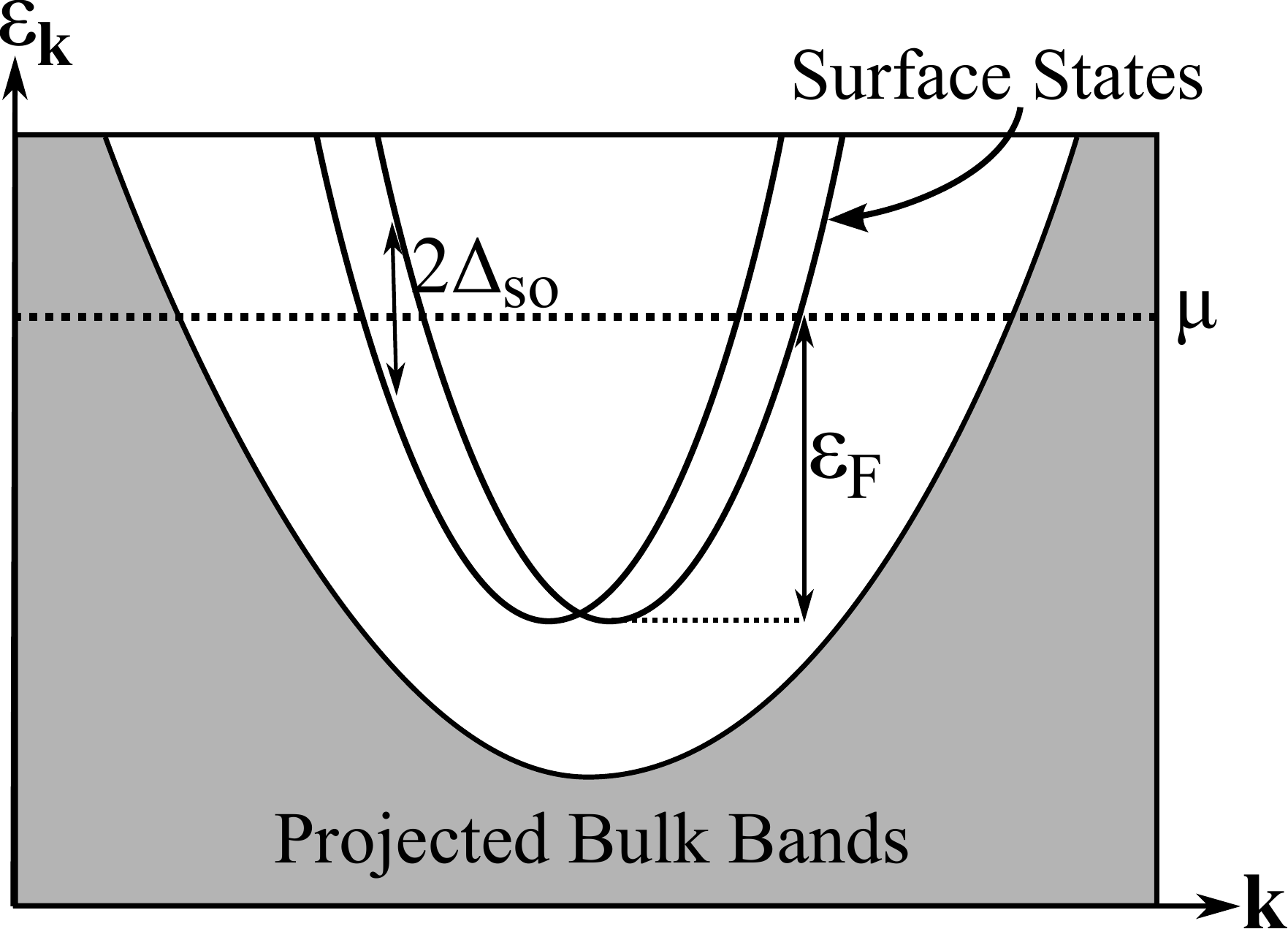}
\end{center}
\vspace{-.2in}
\caption{(a) Simplest possible version of the proposed setup: a strip of Au(111) thin-film (or any other metal with Rashba-split surface state) is deposited on top a conventional superconductor.  An external field is applied parallel to the wire, in order to drive the system into a topological SC state.  Majorana end-states can be detected by tunneling, e.g. with an STM tip. (b) So long as the surface--state survives the deposition of a gate--dielectric, the surface state chemical potential can be controlled by a top--gate. (c) Tunneling density of states, $N(\e)$, as a function of energy, $\e$; the full superconducting gap $\Delta_{\text{B}}$ is induced on the bulk states, by proximity effect.  The surface gap develops a smaller gap $\Delta_{\text{S}}$ due to indirect scattering from disorder and interactions.  (d) Sketch of band-structure of metal with a Rashba spin-orbit split surface band.  Bulk states are projected onto the plane of the surface, and non-zero bulk projected density of states is indicated by gray shading.  The surface-state band forms within a region momentum space where there is no bulk states.  The figure shows a one-dimensional cut through the surface Brillouin zone.  The chemical potential, $\mu$, is represented by a dashed line.  The surface Fermi-energy, $\e_F$, and spin-orbit splitting at the Fermi--surface, $\Delta_{\text{so}}$, are indicated for the surface bands.}
\label{fig:GeometryTunnelingBands}
\vspace{-.1in}
\end{figure*}

The problems associated with small spin--orbit coupling led us to propose building a topological superconductor metallic surface--states\cite{ACPDisorder,ACPMultiBand1,ACPMultiBand2}.
Surface states of heavy-metals are typically tightly bound to the surface, with very small spatial extent.  Consequently, the surface--state electrons are strongly effected by the inversion asymmetry of the surface--interface, generating large Rashba spin-orbit couplings.  For example, the Au(111) surface hosts a well studied surface--state with Rashba splitting of $\Delta_{\text{so}}\approx 50$meV\cite{Au111Surface}, orders of magnitude stronger than the best $\Delta_{\text{so}}$ available in semiconductor nanowires.  Even larger Rashba splittings, $\Delta_{\text{so}}\approx 0.5$eV, are available in the surface states of the Ag(111) surface alloyed with Bi and Pb\cite{AstSurfaceAlloy}. 

In this paper, we develop this proposal in greater detail.  The proposed setup is shown in Fig. \ref{fig:GeometryTunnelingBands}a. A thin metallic film is deposited on top of a convention superconductor.  By the proximity effect, the bulk states of the metal film will inherit some of the superconducting gap $\Delta_0$ from the nearby superconductor.  If the metal film thickness is smaller than or comparable to the superconducting coherence length, $\xi_0$, then the induced bulk gap, $\Delta_B$, will be large ($\Delta_B\approx\Delta_0$).  However, the surface state on the top surface of the metal is nominally isolated from the bulk states and does not couple directly to the superconductor (see the Fig. \ref{fig:GeometryTunnelingBands}d).  Instead, we must rely on disorder and interactions to provide some mixing between the surface--state and bulk bands in order to transmit the bulk SC to the surface--state.  Because SC develops in the surface--state only through indirect scattering processes, the surface--pairing gap, $\Delta_S$ will generically be smaller than the bulk pairing gap, $\Delta_B\approx \Delta_0$.  In this case, the surface--state SC can be revealed by tunneling measurements, which will show a coherence peaks at the edge of the bulk gap, and a smaller sub-gap corresponding to $\Delta_S$ (see Fig. \ref{fig:GeometryTunnelingBands}c).

Once SC is established, one can pattern the metallic film into a quasi-one dimensional wire.  By applying a magnetic field, one can remove the sub-band degeneracy, and tune the chemical potential so that an odd-number of sub-bands is occupied.  If the width of the wire is comparable or smaller than $\xi_0$, then, occupying an odd number of sub-bands will result in Majorana end-states protected by the surface-state pairing gap $\Delta_S$\cite{ACPMultiBand1,ACPMultiBand2,DasSarmaMultiband,BrouewerMultiband}.  

In the simplest version of the proposed setup, shown in Fig. \ref{fig:GatingGeometry}a, tuning to an odd number of sub-bands is accomplished simply by applying an external magnetic field, without gating.  The simplicity of this setup, consisting just of a metallic strip on a bulk superconductor, makes it promising for the initial detection of Majoranas.  To perform more complicated experiments, in which Majoranas are braided, it is necessary to control the local topological phase of different segments of the wire.  For this purpose, one could also add a top gate, as shown in \ref{fig:GeometryTunnelingBands}b.  In order for the top-gate geometry to work, one needs to check that the surface--state is not destroyed by the presence of the gate dielectric.

While the original proposals for creating topological SC from Rashba SOC required applying a field perpendicular to the surface--plane\cite{SauSemiconductorSpinOrbit}, it has since been pointed out\cite{LutchynWires,OregWires} that once the electron motion is confined along a quasi-one-dimensional wire, a parallel field along the wire can also create topological SC. Previous discussions also emphasized that the parallel field configuration is advantageous because, compared to the perpendicular field, the adjacent superconducting film will be less affected by harmful orbital effects\cite{LutchynWires,OregWires}.  Here we point out a further advantage of the parallel field setup for multi-channel wires: so long as the Zeeman splitting--energy from the field $B$ is larger than $\Delta_S$, then the wire will exhibit Majorana end-states for arbitrarily large chemical potential $\mu$.  This is in marked contrast to the perpendicular field case, in which topologically non-trivial regions were only available for a small range chemical potentials near the Rashba-band crossing.  The ability to operate at arbitrarily large chemical potential frees one from fine-tuning the chemical potential near the Rashba crossing, and allows one to work at much larger carrier-densities and spin-orbit couplings (since $\Delta_{\text{so}}$ grows with the Fermi-momentum).

The outline of the paper is as follows:  we begin by demonstrating that using a parallel field allows one to achieve topological SC and Majorana end-states at arbitrarily large chemical potentials in multi-band wires. Next, we describe how SC can be induced on the surface--state by impurity scattering and interactions.  Our analysis suggests that it may be advantageous to artificially disorder the surface in order to enhance $\Delta_S$.  We then confront an often voiced concern\cite{GatingConcern1,GatingConcern2}, that the presence of the nearby metal (superconductor) will make it impossible to control the chemical potential of the surface--states (nanowires) respectively.  This concern has led to some rather complicated proposals that attempt to avoid the perceived gating problem\cite{GatingConcern1,GatingConcern2}.  Here we show that, under realistic experimental conditions, it should not actually be difficult to tune the surface--state chemical potential over a wide range of $\approx$ 100meV.  Therefore, conventional top-gates should be sufficient to tune the wire into a topological state, and to manipulate Majoranas.  Finally, we discuss a particularly promising candidate material, the Au(111) surface.

We believe that the simplicity and robustness of the proposed setup make it a very promising route to realizing Majorana fermions.  The large spin--orbit coupling available in metallic surface--states allow for larger intrinsic superconducting gaps, and render the topological SC effectively immune to disorder. 

\begin{figure*}[ttt]
\begin{center}
{\bf a)}\hspace{-.2in}\includegraphics[height=2in]{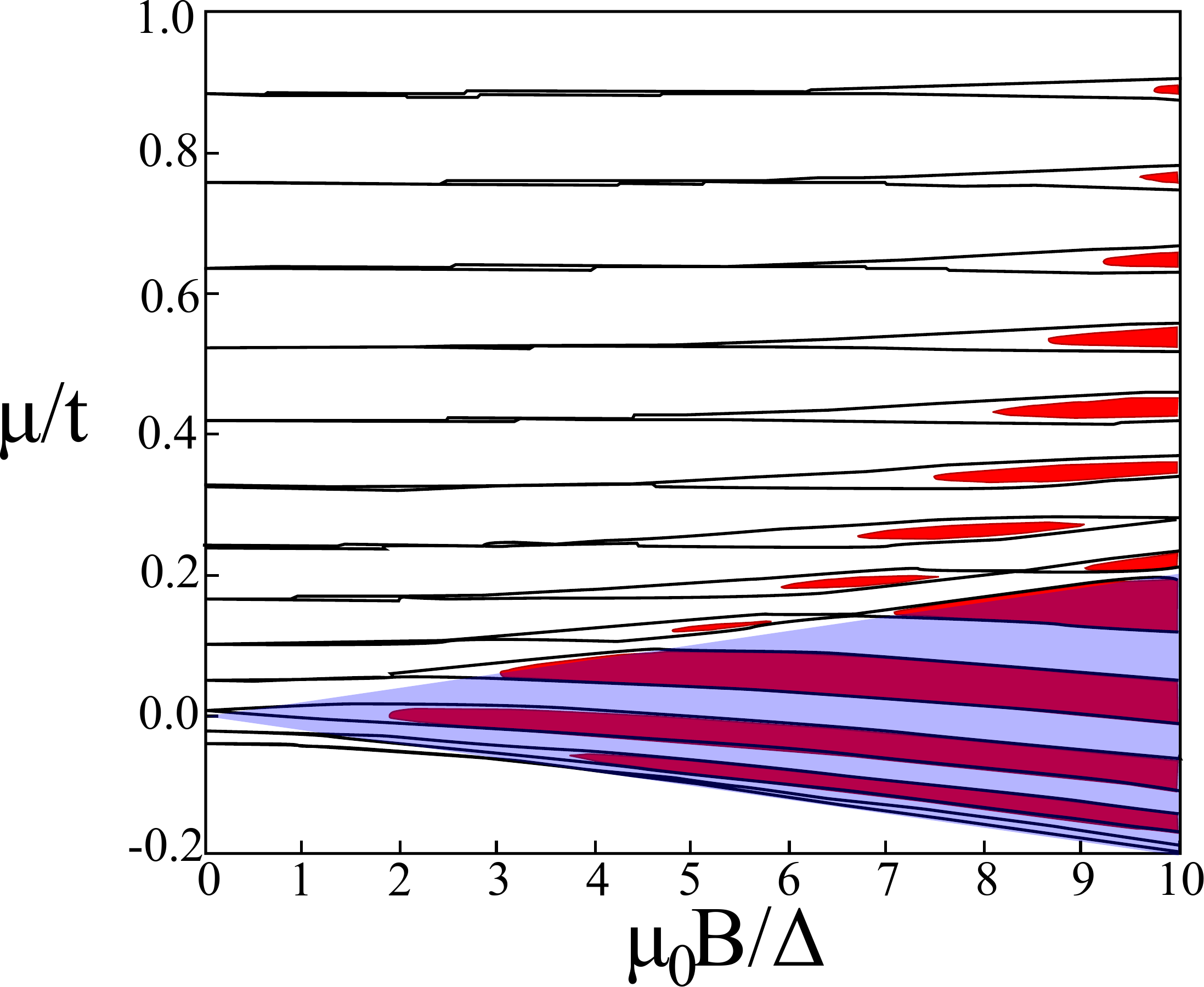}
\hspace{.2in}{\bf b)}\hspace{-.2in}
\includegraphics[height=2in]{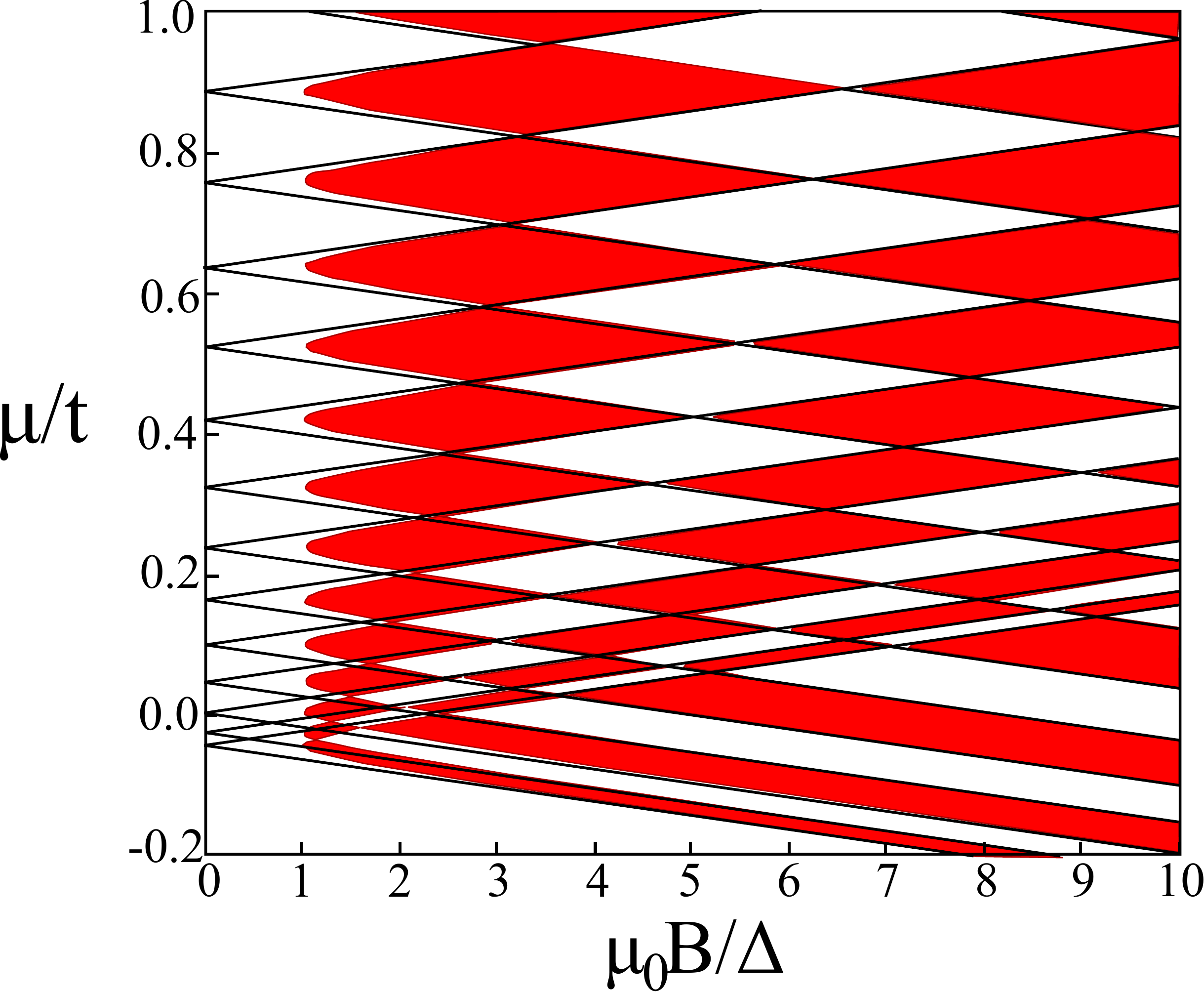}
\hspace{.2in}{\bf c)}\hspace{-.3in} \includegraphics[height=2.1in]{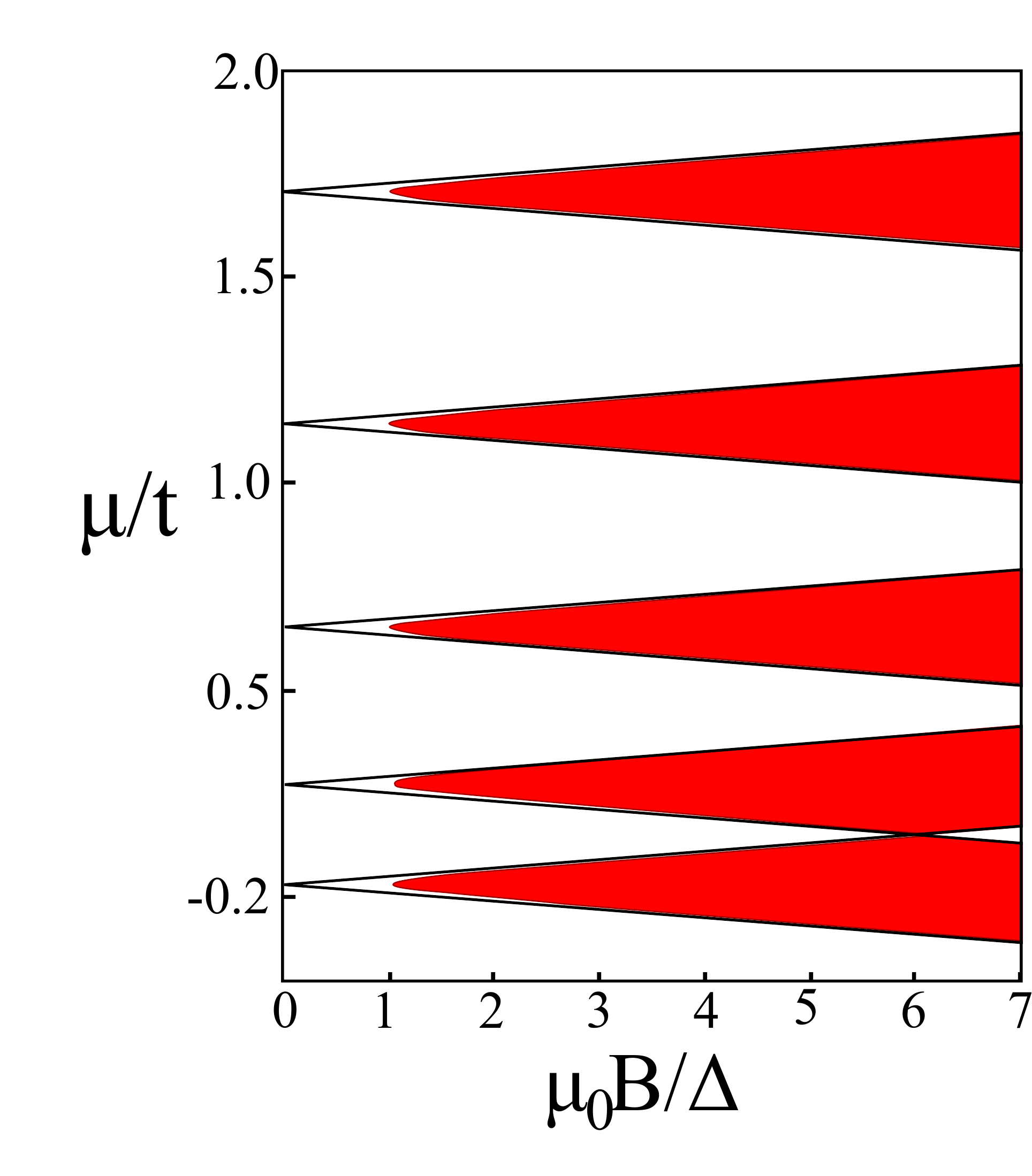}
\end{center}
\vspace{-.2in}
\caption{(a) Numerical phase diagram for 40-site wide wire in perpendicular field ($\v{B}\sim\hat{z}$), as a function of chemical potential $\mu$ and magnetic field $\mu_0B$. Black lines indicate sub-band bottoms in the normal state (without superconductivity), red filled regions indicate the presence of Majorana end-states, which occur when the sub-band degeneracy is removed and the sub-band splitting is sufficiently larger than the pairing gap $\Delta$.  The sub-bands are initially degenerate for $B=0$, and split quadratically as $B$ is increased.  Blue shaded region indicates the Zeeman gap for a full two-dimensional sample.  In the wire, the topological region extends slightly outside the Zeeman gap for sufficiently large $B$ (for $\mu_0B^2/\Delta_{SO}\gtrsim\Delta)$.  Simulation parameters: $t=50$, $\alpha_R^2/t=10$, $\Delta_S=1$.
(b) Same setup described in a) but with the magnetic field applied along the wire ($\v{B}\sim\hat{x}$).  Unlike the perpendicular field case, the wire always remains in the topological region so long as $\mu_0B>\Delta_S$ and an odd number of sub-bands is occupied.  Unlike the parallel field case, the black lines that indicate sub-band bottoms split linearly in the applied field, giving rise to a criss-crossing diamond pattern of topological and non-topological phases. (c) Parallel field phase diagram for 10-site wide wire, topological regions occupy smaller fraction of the phase-diagram.}
\label{fig:TopologicalPhaseDiagram}
\vspace{-.1in}
\end{figure*}

\section{Sub-band Spectrum and Topological Phase Diagram}
The Bugoliobov-de-Gennes Hamiltonian for the system with Rashba SOC, Zeeman splitting, and induced SC is:
\begin{align} \label{eq:FullHamiltonian} H_{\v{k}} = \[\xi_k+\alpha_R \hat{z}\cdot\(\boldsymbol{\sigma}\times\v{k}\)\]\tau^3- \mu_0\v{B}\cdot\boldsymbol{\sigma}-\Delta_S\tau^1
\end{align}
where $\{\boldsymbol{\sigma}\}$ are the spin-Pauli matrices, and $\{\boldsymbol{\tau}\}$ are Pauli matrices in the BCS particle--hole basis.  Here $\xi_k = \frac{k^2}{2m}-\mu$ is the spin-indenpendent part of the dispersion, $\alpha_R$ is the Rashba velocity related to the spin-orbit coupling by $\Delta_{\text{so}}=\alpha_Rk_F$ (where $k_F$ is the Fermi-momentum), $\v{B}$ is the magnetic field which couples to the spin with the effective magnetic moment $\mu_0 = g\mu_B$, and $\Delta_S$ is the SC gap.  The chemical potential $\mu$, is measured with respect to the Rashba-band crossing in the absence of $B$ and $\Delta_S$.

Furthermore, we consider electrons confined to a quasi-one-dimensional strip of width $W$ along the y-direction and length $L$ along the x-direction.  The wire will exhibit Majorana end-states under certain conditions, which are outlined below.

Topologically non-trivial states arise only when 1) the splitting between adjacent sub-bands is larger than the pairing gap $\Delta_S$ and 2) an odd number of transverse sub-bands is occupied.  Condition 1) ensures that pairing does not mix in states from neighboring sub-bands strongly enough to drive the system into the topologically trivial state.  Furthermore, it allows one to meaningfully speak of the ``number of occupied sub-bands", even though strictly speaking, this concept is well defined only the absence of pairing.  Condition 2) ensures that there are an odd number of Majorana end-states (one for each channel), which is guaranteed to leave one decoupled Majorana mode at zero-energy.  We emphasize, that while condition 2) is stated explicitly in terms of number of sub-bands, the structure of the topological phase diagram is qualitatively similer even for smoothly meandering wires or wires with smooth spatial varations in widths for which sub-bands are not well defined\cite{ACPMultiBand2}.

In the absence of the Zeeman field, $\mu_0B=0$, time-reversal symmetry is intact and sub-bands occur in pairs.  Generically, without breaking time-reversal symmetry it is impossible to occupy an odd number of sub-bands.  Applying a magnetic field perpendicular to the electron spins perpendicular to the plane lifts the degeneracy and splits the energy spectrum into a series of individual (non-degenerate) sub-bands.  If the sub-band splitting is sufficiently large compared to the induced superconducting pairing, then it is possible to drive the system into a toplogical non-trivial state by tuning the magnetic field or chemical potential.

For two-dimensional Rashba systems, the only way to achieve a chiral topological superconductor is to apply a Zeeman field perpendicular to the plane, and tune the chemical potential within the Zeeman gap.  For small spin-orbit coupling, the Zeeman gap occurs at low energy, forcing one to operate at low carrier densities and small energy scales. By contrast, in a quasi-one-dimensional wire the electron motion occurs predominantly along the wire, and consequently due to the Rashba spin-orbit coupling, the electron spins point predominantly perpendicular to the wire but in the plane.  Unlike in 2D, applying a Zeeman-field along a wire also serves to split the sub-band degeneracies.  We will see below, that applying a field parallel to the wire allows one to operate at arbitrarily large chemical potential, well outside the regime in which the bulk 2D system would be topologically non-trivial.

\subsection{Out-of-plane Field}
The dispersion without superconductivity in the presence of a perpendicular field is:
\begin{align}\label{eq:DispersionPerpField} \e_{\v k,\lambda} = \frac{k^2}{2m}-\mu+\lambda\sqrt{\alpha_R^2k^2+(\mu_0B)^2}
\end{align}
where $\lambda = \pm 1$.  The resulting phase diagram for a superconducting wire obtained from numerical simulation is shown in Fig. \ref{fig:TopologicalPhaseDiagram}a. Topological phase transitions occur when the chemical potential coincides with the bottom of a transverse sub-band, so long as the transverse sub-band spacing is larger than $\Delta_S$. For $\mu>0$,\cite{SubbandEndnote} the sub-band splitting due to the applied field can be estimated by setting $k_x=0$ and $k_y\approx \pm k_F$ in Eq. \ref{eq:DispersionPerpField}.

Consider the energy $\e_n$ of the $n^{\text{th}}$ sub-band for $B=0$.  For $k_x=0$ there are four different states with $\e_{\lambda,(k_x=0,k_y)}=\e_n$, labeled by different $k_y$.  In the wire, linear superpositions of these four states are formed to satisfy the hard-wall boundary-conditions (which can only be satisfied for a discrete set of energy values).  In the absence of a magnetic field, the four $k_y$ states at energy $\e_n$ form two degenerate combinations related by time-reversal symmetry.  Due to the Rashba SOC, the spin of each of the $k_y$ states lies in the plane, and a perpendicular field does not directly mix the two states.  Consequently, the sub-band splitting from the field  occurs through virtual admixture of higher energy states, and scales
like $\Delta E_{\text{sb}} \approx \frac{B^2}{\Delta_{\text{so}}}$ where $\Delta_{\text{so}}=\alpha_Rk_F$.  Inside the bulk Zeeman gap, ($|\mu|<\mu_0B$), it is always possible to occupy an odd number of sub-bands.  As chemical potential is increased outside of the bulk Zeeman-gap, $\Delta_{\text{so}}$ increases until $\Delta E_{\text{sb}} < \Delta_S$, at which point the topologically nontrivial regions stop occuring.

\subsection{Field along the Wire}
The dispersion without superconductivity in the presence of a parallel field is:
\begin{align}\label{eq:DispersionParallelField} \e_{\v k,\lambda} = \xi_k+\lambda\alpha_R\sqrt{k_x^2+\(k_y+\frac{\mu_0B}{\alpha_R}\)^2}
\end{align}
For $\mu>0$,\cite{SubbandEndnote} the sub-band bottoms occur for $k_x=0$.  Unlike the perpendicular field case described above, the initially degenerate sub-band bottoms are split linearly by parallel $B$, $\Delta E_\text{sb}\approx \mu_0B$, independent of the spin-orbit coupling strength.  The linear sub-band Zeeman--splitting leads to the criss-crossing pattern of diamonds shown in Fig.\ref{fig:TopologicalPhaseDiagram}b.  So long as $\mu_0B>\Delta_S$, we expect to be able to occupy an odd number of sub-bands and achieve a topologically non-trivial state with Majorana end-states.  The topological phase diagram obtained from numerical simulations and shown in Fig. \ref{fig:TopologicalPhaseDiagram}b,c bears out this expectation, exhibiting topologically non-trivial phases for arbitrarily large chemical potential.  

Two illustrative cases are shown in Fig. \ref{fig:TopologicalPhaseDiagram}b,c. In Fig. \ref{fig:TopologicalPhaseDiagram}b the sub-band spacing is comparable to the SC gap, $\Delta E_{\text{sb}}\approx\Delta_S$, corresponding to the metallic strip having width comparable to the SC coherence length, $W\approx \xi_0$.  In this case, the topological and non-topological phases occupy roughly equal portions of the phase diagram, allowing one to more easily tune into the topological region by changing $B$ or gate voltage.  In Fig. \ref{fig:TopologicalPhaseDiagram}c the sub-band spacing is larger than the SC gap, $\Delta E_{\text{sb}}>\Delta_S$, corresponding to $W<\xi_0$.  Here, for small fields, the non-topological regions occupy a larger fraction of the phase diagram.  

Having $\Delta E_{\text{sb}}\approx\Delta_S$ is especially important for the gateless setup shown in Fig.\ref{fig:GeometryTunnelingBands}a, where tuning sub-band number is accomplished purely by changing $\mu_0B$. If $\Delta E_{\text{sb}}\gg\Delta_S$, then, without controlling $\mu$, the wire is most likely to be deep in the topologically trivial region.  This would likely require applying large $\mu_0B\gg\Delta_S$ in order to tune into the topological phase.  In contrast, for $\Delta E_{\text{sb}}\approx\Delta_S$, the maximum require $\mu_0B$ is $\approx\Delta_S$ regardless of the initial $\mu$, allowing one to readily tune to the topological phase without controlling $\mu$.

\section{Indirectly Induced Surface--State Superconductivity}
Having described the advantages of applying a magnetic field along the wire, we now address the issue of how superconductivity is induced on the surface--state.

Consider a thin film of a spin-orbit coupled metal with a surface state, deposited on top of a conventional s-wave superconductor.  If the metal is in good contact with the superconductor and the film thickness does not greatly exceed the superconducting coherence length, $\xi_0$, then nearly the full superconducting gap $\Delta_B\approx\Delta_0$ will be induced in the bulk-bands of the metal film.  

However, in a pristine sample and in the absence of interactions, the metal surface-state has no overlap with the bulk metal bands (see Fig.\ref{fig:GeometryTunnelingBands}d).  Consequently we must rely on indirect scattering between the bulk and surface bands to transmit the bulk superconductvity to the surface states.  This indirect scattering can occur either by elastic scattering off of static impurities, or by inelastic scattering due to Coulomb interactions or phonons.  Below we discuss both types of scattering, starting with the simpler case of elastic impurity scattering.

\subsection{Surface-Bulk Mixing from Elastic Impurity Scattering}
As a simple model of screened impurities, we consider a random potential $V_{\text{imp}}(\v{r})$ with zero average $\overline{V_\text{imp}(\v{r})}=0$ and short range correlations, $\overline{V_{\text{imp}}(\v{r})V_{\text{imp}}(\v{r}')}=W^2\delta(\v{r}-\v{r'})$.  Here $\overline{(\dots)}$ indicates averaging over impurity configurations.  The impurity scattering from the surface state to the bulk bands gives rise to the following self-energy, evaluated within the self-consistent Born approximation:
\begin{align}\label{eq:ImpuritySelfEnergy} \Sigma_{\text{imp}}(i\omega) &= \overline{V_{\text{imp}}(\v{r}_\parallel,z=0)G_B(\v{r},0;\v{r'}_\parallel,0)V_{\text{imp}}(\v{r}_\parallel,z=0)} \nonumber \\ &= -W^2\tau_3\sum_k  \frac{i\omega-\Delta_0\tau_1}{\omega^2+\xi_k^2+\Delta_B^2}\tau_3 
\nonumber\\
&= -\frac{1}{2\tau_B}\frac{i\omega-\Delta_0\tau_1}{\sqrt{\omega^2+\Delta_B^2}}
\end{align}
Here $\omega$ is the Matsubara frequency corresponding to Fourier transforming in imaginary time, $\tau_B = 2\pi\nu_B(0)W^2$ is the elastic scattering time for bulk electrons, $\nu_B(0)$ is the density of states at the bulk Fermi-surface, and $G_B = 1/\(i\omega-\xi_k\tau_3-\Delta_B \tau_1\)$ is the Green's function for bulk fermions with dispersion $\xi_k$ and bulk pairing gap $\Delta_B$.  This expression is valid so long as localization corrections can be ignored in the bulk, i.e. so long as $\e_{F,B}\tau_B\gg 1$, where $\e_{F,B}$ is the bulk Fermi-energy.  We emphasize that the impurity induced surface--bulk mixing is sensitive only to impurities near the surface.

Incorporating $\Sigma_{\text{imp}}$ into the surface state Green's function yields:
\begin{align} G_S(i\omega,\v{k}) &= \[\(G_S^{(0)}\)^{-1}-\Sigma_{\text{imp}}\]^{-1} 
\nonumber\\ &= \frac{Z(i\omega)}{i\omega-Z(i\omega)H_0-(1-Z(i\omega))\Delta_B\tau_1} \end{align}
where $G_S^{(0)}(i\omega) = \[i\omega-H_0\]^{-1}$ is the bare surface Green's function, $H_0 = (\frac{k^2}{2m}-\mu-\alpha_R\hat{z}\cdot\(\boldsymbol{\sigma}\times\v{k}\))\tau_3-\mu_0B\sigma_z$ is the surface Hamiltonian,  and 
\begin{align} Z(i\omega) = \[1+\frac{1/2\tau_B}{\sqrt{\Delta_B^2+\omega^2}}\]^{-1} \end{align}
is the surface quasi-particle residue.

The effective pairing gap from impurity induced surface--bulk mixing is given by smallest pole of $G_S$ which occurs at frequency $\omega_p$ defined by: $\(\frac{\Delta_B}{\omega_p}-1\)^2=4\tau_B^2\(\Delta_0^2-\omega_p^2\)$.  For the limiting cases of strong and weak disorder the induced gap reads:
\begin{align} \label{eq:ImpurityInducedGap} \Delta_{\text{imp}} = \omega_p = \left\{\begin{array}{ll} (1-4\Delta_B^2\tau_B^2)\Delta_B ;& \Delta_B\tau_B\ll 1 \\ 1/2\tau_B ;& \Delta_B\tau_B\gg 1\end{array}\right.\end{align}
For strong disorder, $\Delta_B\tau_B\ll 1$, the induced gap is nearly equal to the full bulk gap, whereas for weak disorder, $\Delta_B\tau_B\gg 1$ only a small fraction of the bulk gap is transmitted to the surface state.

Eq. \ref{eq:ImpurityInducedGap} suggests that if the surface states are too well isolated from the bulk bands, then it may actually be advantageous to introduce surface disorder to ensure sufficient mixing of the surface and bulk bands. However, in order to drive the system into a topological superconducting state one must apply an external magnetic field, in which case time-reversal symmetry is broken and disorder is pair-breaking\cite{ACPDisorder,BrouwerDisorder,StanescuDisorder}. One might therefore worry that increasing disorder may tend to suppress rather than enhance superconductivity. However, the size of the pair-breaking component of disorder scattering was shown to be strongly dependent on the ratio of the spin-orbit coupling $\Delta_{\text{so}}$ to the Zeeman splitting $\mu_0B$\cite{ACPDisorder}.  In particular, for very strong spin-orbit coupling, the pair--breaking effects of impurities is small.
\begin{figure}[bbb]
\vspace{-.1in}
\begin{center}
\includegraphics[width=3.2in]{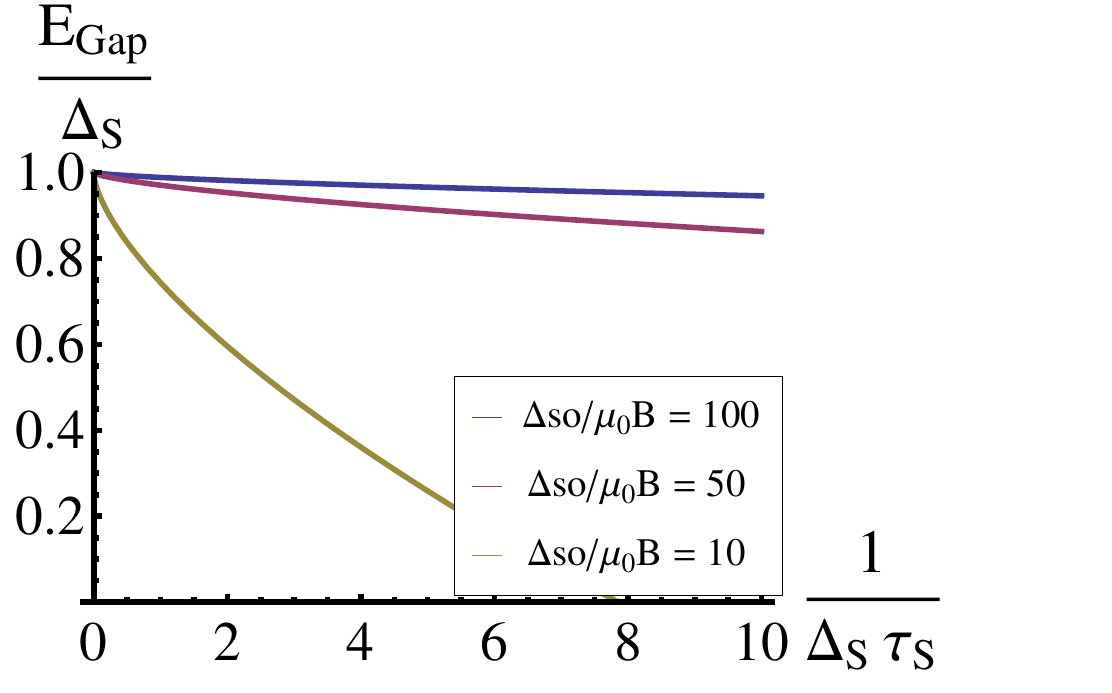}
\end{center}
\vspace{-.1in}
\caption{Surface pairing gap, $\Delta_S$, for various $\Delta_{\text{so}}/\mu_0B$.  The reduction of the induced surface--gap due to disorder is very weak for $\Delta_{\text{so}}\gg \mu_0B$.  Plot is generated from the calculations of disorder induced pair breaking from Ref. \onlinecite{ACPDisorder}.}
\vspace{.3in}
\label{fig:DisorderedGap}
\end{figure}

For heavy metal materials with surface states $\Delta_{\text{so}}$ is commonly quite large, on the order of $\approx 100$meV\cite{Au111Surface,AstSurfaceAlloy,PbSurface}.  In contrast, the typical Zeeman splitting needed is of the order $\mu_0B\approx 2\Delta_0\approx 1$meV.  In this regime, the reduction of the surface pairing gap, $\delta\Delta_S$, due to disorder will be quite small\cite{ACPDisorder}: 
\begin{align} \(\frac{\delta\Delta_S}{\Delta_S}\)_\text{disorder} &\approx -\(\frac{(\mu_0B)^2}{\Delta_{\text{so}}^2}\frac{1}{\Delta_S\tau_S}\)^{2/3}
\nonumber \\
&\approx -10^{-3}\(\frac{1}{\Delta_S\tau_S}\)^{2/3}
\end{align}
where $\tau_S$ is the elastic lifetime for surface-states due to disorder.  The reduction of the SC gap due to surface disorder is shown in Fig. \ref{fig:DisorderedGap}, where for $\Delta_{\text{so}}/\mu_0B\approx 100$ we see almost no effect at all from disorder.  Therefore, so long as spin-orbit coupling is large, it is possible to enhance the surface--state pairing by adding disorder without suppressing the pairing gap by pair-breaking scattering.

\subsection{Surface-Bulk Mixing from Inelastic Scattering}
\begin{figure}[ttt]
\begin{center}
\includegraphics[width=2.85in]{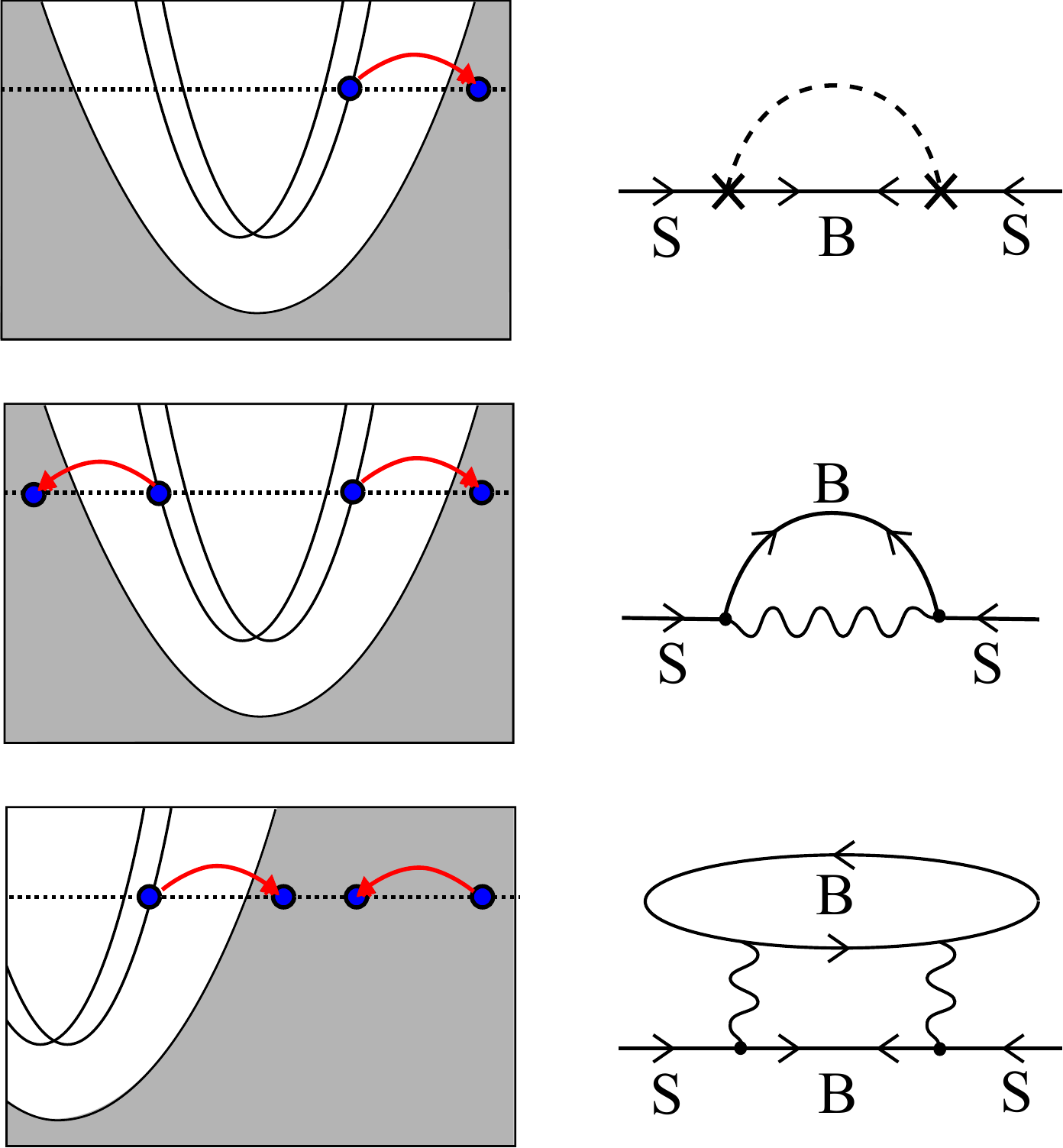}
\end{center}
\vspace{-.2in}
\caption{Depiction of virtual scattering processes which mix bulk and surface bands and generate surface superconductivity (left column) along with representative Feynman--diagrams (right column).  In the diagrams, lines labeled by 'S' and 'B' indicate surface--state and bulk--state propagators respectively; propagators with left (right) arrows are conventional particle (hole) propagators, whereas propagators with both left and right arrows are anomalous propagators due to the Cooper-pair condensate.  Each process shown in the left column represents half of the corresponding diagram (to complete the diagram, the process is repeated in reverse).  The top row depicts elastic scattering from impurities, represented diagrammatically by $\times$'s connected by a dashed line (indicating scattering off of the same impurity).  The middle and bottom rows show inelastic processes that generate surface pairing; wavy-lines represent either screened Coulomb interactions or phonons.  The middle row shows inelastic pair-scattering from surface--to--bulk, and the bottom row shows interaction induced surface--bulk tunneling which is accompanied by the creation of a bulk particle--hole pair.}
\label{fig:ScatteringProcesses}
\vspace{-.1in}
\end{figure}
The surface-state and bulk bands are also mixed by inelastic electron-electron scattering and electron-phonon scattering.  The middle and bottom rows of Fig. \ref{fig:ScatteringProcesses} illustrate two processes that induce pairing in the surface state.  In the process shown in the middle row, a pair of surface--electrons are virtually scattered into bulk states, where they develop pair correlations before returning to the surface.  The process shown in the bottom row shows interaction driven (virtual) tunneling between surface and bulk states accompanied by a virtual particle-hole excitation.

In contrast to the surface--bulk mixing, which depends only on the easily measurable quantities $\tau_B$ and $\Delta_B$, the inelastic surface--bulk mixing is difficult to accurately estimate.  Doing so would require detailed knowledge of screening properties, phonon dispersion, and electron-phonon coupling matrix elements.  These quantities are highly non-universal, and difficult to measure. Therefore, rather than attempting a detailed calculation, we simply illustrate that interaction driven processes can also contribute to surface--state superconductivity.

\section{Gating Metallic Surface States}
One often stated worry about proposals to realize Majoranas in nanowires with induced superconductivity, is that, since the wire is necessarily in good contact with a superconductor, the chemical potential of the wire may be pinned to the Fermi-energy of the superconductor making it impossible to gate the nanowire.  This worry would also apply to the setup discussed here, using metallic surface states.

Here we address this worry, and demonstrate that the pinning of the surface chemical potential due to the bulk Fermi-surface is not strong enough to prevent gating.  Rather, under experimentally realistic assumptions it should be straightforward to tune the surface chemical potential across 100's of sub-bands.  

\begin{figure}[ttt]
\vspace{.1in}
\begin{center}
\includegraphics[width=2.5in]{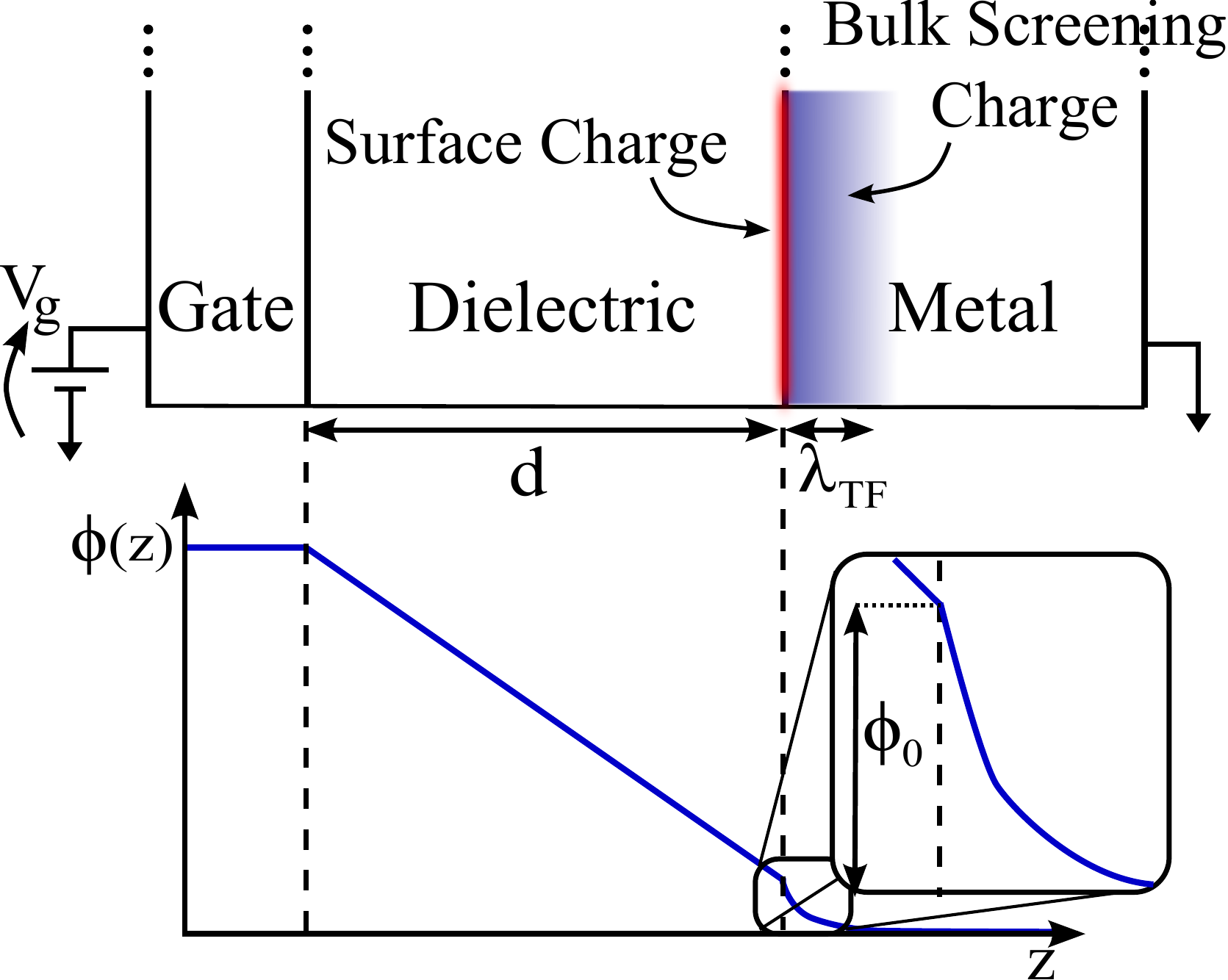}
\end{center}
\vspace{-.3in}
\caption{Electrostatic potential profile from applied gate voltage(bottom) aligned with the proposed materials stack (top, shown here rotated $90^\circ$ relative to Fig. \ref{fig:GeometryTunnelingBands}b).  The surface chemical potential is shifted by $\delta\mu_s=-e\phi(0)$ relative to the bulk chemical potential.  Estimates using typical material parameters demonstrate that one can readily tune the chemical potential by $\pm 100$meV, despite the presence of a large density of states from the metallic bulk.}
\vspace{.3in}
\label{fig:GatingGeometry}
\end{figure}

Consider applying a voltage, $V_g$, to a gate separated from the surface of the grounded metal sample by a dielectric of dielectric constant $\epsilon$ and thickness $d$ (see Fig. \ref{fig:GatingGeometry}).  The applied voltage induces a bulk screening charge density $\rho_B(z)$ confined within a screening length, $\lambda_{TF}= \sqrt{\frac{\epsilon_0}{e^2N_B}}$, of the surface, and also induces a surface--state charge density $\rho_s$.
For simplicity, we assume that the extension of the surface-state into the bulk is much smaller than the screening length $\lambda_{TF}= \sqrt{\frac{\epsilon_0}{e^2N_B}}$, and approximate the surface state as infinitesimally thin.  Incorporating a finite surface--state width is straightforward, but does not substantially alter the results.

Within the Thomas-Fermi approximation the bulk screening charge is: $\rho_B(z) = -e^2N_B\phi(z)$ where $N_B$ is the bulk density of states, and the induced surface charge is $\rho_s(z) = -e^2N_s\phi(0)\delta(z)$, where $N_s$ is the surface density of states, and $\phi(0)$ is the chemical potential at the metal surface ($z=0$).  Solving Poisson's equation we find for the surface potential
\begin{align} \label{eq:Phi0}
\phi(0) = \epsilon_R\frac{\lambda_{TF}}{1+{N_S}/{\lambda_{TF}N_B}}\frac{V_g}{d}\end{align}
where $\epsilon_R$ is the relative permittivity of the gate dielectric.

We see that the consequence of applying the gate voltage is to shift the chemical potential of the surface by $\delta\mu_S=-e\phi(0)$ compared to the bulk chemical potential.  For typical metals, $v_F\approx 1\times 10^6$m/s, and the bulk and surface band masses are comparable to the bare electron mass, giving $\frac{N_S}{\lambda_{TF}N_B}\approx 4$.  Break-down fields for typical gate dielectrics (e.g. $\text{SiO}_2$) are of on the order of $E_{\text{max}}\approx 1$V/nm, and typical screening metallic lengths are $\lambda_{\text{TF}}\approx 1\AA$.  For SiO$_2$, with $\epsilon_R=4.9$, this gives $\delta\mu_s^{(\text{max})}\approx \pm 100$meV.  In comparison, for a metallic wire with width of the order of the superconducting coherence length, the typical sub-band spacing is $\approx \Delta_0 \approx 1$meV, indicating that one could tune across hundreds of sub-bands.  Furthermore, using a higher-K dielectric such as HfO$_2$ would allow one to tune the surface-potential over an even larger range.

From simple electrostatic modeling, we have shown that the close proximity to a metal does not substantially impede the ability to tune the surface-state chemical potential by a gate voltage.  This analysis also implies that one could use a top gate to control the chemical potential of semiconducting nanowires placed on top of a superconductor.  However, in order to get strong proximity induced SC, it is typically necessary to deposite nanowires on an insulating substrate and coat them with a superconductor.  In this setup, one would need to employ a back-gate, which offers poor electrostatic control (since the wire would be coated on three sides by superconductor).  Therefore, more complicated geometries are required; for example, one could partially coat the wire with superconductor and partially with a gate\cite{Akhmerov}. In contrast, the metallic-surface state chemical potential can be tuned using just a simple top-gate geometry, substantially simplifying the fabrication requirements.

\section{Mini--Gap}
While the gateless geometry of Fig. \ref{fig:GeometryTunnelingBands}a. is very simple, there are advantages to the top-gate geometry shown in Fig. \ref{fig:GeometryTunnelingBands}b.  For example, it has been shown that, in the presence of multiply occupied sub-bands, the Majorana zero modes are accompanied by sub-gap fermion states localized at the wire--ends\cite{ACPMultiBand1}.  These localized fermions have energy spacing on the order of the so-called ``mini-gap" $\Delta_{\text{mg}}<\Delta_S$.  Recently, it was shown that the maximal mini-gap spacing occured when the wire-width was comparable to the superconducting coherence length, and for perfect rectangular strips\cite{MiniGapScaling}, the optimal minigap scales as $\Delta_{\text{mg}}\approx \Delta_S^2/\e_F\ll \Delta_S$.\cite{BrouwerMiniGap}  We believe that the scaling $\Delta_{\text{mg}}\approx \Delta_S^2/\e_F$ is partially an artifact of the assumption of a perfectly rectangular geometry, which leads to Majorana end-states for each band that are almost perfectly orthogonal, and therefore only very weakly mixed.  For the more realistic case, where the wire-end is rounded (or otherwise distorted) on length-scales $\approx 1/k_F$, then the end-states have randomized overlaps, leading to a slightly more favorable mini-gap scaling that should be of the order $\Delta_{\text{mg}}\approx \Delta_S\sqrt{\Delta_s/\e_F}$ (see Appendix A. below).  For the case of Au, we have $\e_F\approx 0.5$eV and optimistically one could use a large gap superconductor such as Nb so that $\Delta_S\approx 1$meV, giving $\Delta_{\text{mg}}\approx 200$mK, which is potentially resolvable in a dilution refrigerator. 

While these minigap states are known not to disrupt topological operations involving spatially well separated Majoranas\cite{AkhmerovQubit}, the small mini-gap states complicate tunneling based probes of the Majorana zero-modes unless the temperature and resolution of the probe are lower than $\Delta_{\text{mg}}$.  The presence of a large number of mini-gap states can be avoided by selectively gating sections of the wire so that the local sub-band number changes by at most $\pm 1$.\cite{BrouwerMiniGap}  Here we re-emphasize that this scheme does not rely on the existence of well-defined sub-bands, and that changing the average width by $\pm 1$ sub-bands abruptly will trap a Majorana mode even for meandering wires.

\section{Discussion and Conclusion}
In summary, we believe that metallic thin-films with Rashba split surface states offer a promising route to realizing Majorana fermions.  The large Rashba spin--orbit couplings in these materials offer several advantages over similar proposals involving semiconducting materials, allowing for substantially larger energy scales, and dramatically reducing the sensitivity to disorder.

One particularly promising surface state occurs on the (111) surface of Au\cite{Au111Surface}.  This surface is stable and has been well studied by ARPES. The surface bands have high carrier density, $\e_F\approx 0.5$eV, and large Rashba spin-orbit splitting $\Delta_{\text{so}}\approx 50$meV.  The first task towards creating Majorana fermions in the Au(111) surface state, would be observe the indirectly induced surface pairing gap $\Delta_S$, which could be examined by planar tunneling or STM tunneling (see Fig. \ref{fig:GeometryTunnelingBands}b).  If the surface--bulk band mixing is insufficient to achieve large $\Delta_S$, the surface could be intentionally disordered to improve the surface--pairing.  The measurements involved should be straightforward, and to our knowledge, would constitute the first observation of superconductivity induced onto a surface--state.  

The setup shown in Fig. \ref{fig:GeometryTunnelingBands}a. is the simplest possible version of our proposed scheme.  By patterning a quasi-one-dimensional strip of Au on top of an ordinary superconductor, one can achieve a topological superconductor with Majorana end--states simply by applying a magnetic field parallel to the wire (without ever tuning the chemical potential by gating).  If the wire width is comparable to the coherence length, then only small magnetic fields $\mu_0B\approx \Delta_S$ are required to achieve Majorana end-states.  As a concrete example, taking $\e_F$ and $k_F$ of the Au(111) surface state measured in Ref. \onlinecite{Au111Surface}, and taking $\Delta_S\approx 5K$ gives coherence length: $\xi_0\approx 5\mu$m, corresponds to $n\approx 500$ occupied-subbands.

The Majorana end-states could be detected by tunneling measurements, e.g. by STM or by fabricated tunneling contacts.  Resonant Andreev reflection from a Majorana fermion gives a distinctive quantized conductance: $G = 2e^2/h$.\cite{Law}  As described above, in multichannel wires, the Majorana zero-mode will coexist with other sub-gap states localized to the end of the wire.  These states have energy spacing $\approx \Delta_{mg}$ which is typically $\ll \Delta_S$.  If the mini-gap spacing is too small to experimentally resolve by tunneling, it would still be interesting to show the presence of sub-gap states at the end of a fully gapped superconducting wire.  These sub-gap states would be confined the wire-end and would disappear when $\mu_0B<\Delta_S$ giving a clear signature of topological superconductivity.  We have shown that the parallel field geometry has the advantage of allowing one to operate at arbitrarily large chemical potential.  This observation is important for the Au(111) surface state, because its large $\e_F\approx 0.5$eV could make it difficult to tune the chemical potential near the Rashba crossing (which would be necessary for the perpendicular field setup). 

Finally, we have shown that, in contrast to semiconducting nanowire based proposals, it is possible to effectively control the metallic surface--state chemical potential with a simple top-gate geometry, despite the presence of a large bulk-density of states.  This obviates the need for more complicated gating geometries, or complicated gateless setups such as those proposed in Ref. \onlinecite{GatingConcern1,GatingConcern2}.  This is in contrast to proposals involving semiconducting nanowires, which typically need to be coated with superconductor in order to induce SC by proximity.  For a wire coated with superconductor, a simple top-gate does not exert sufficient electrostatic control over the wire, and more complicated gating geometries are required.

\section*{}
\textit{Acknowledgements --}
We thank J. Moodera and A.R. Akhmerov for helpful conversations.  This
work was supported by DOE Grant No. DE–FG02–03ER46076 (PAL) and NSF IGERT Grant No. DGE-0801525 (ACP).

\newpage 

\appendix
\section{A. Mini-Gap Scaling}
Ref. \onlinecite{BrouwerMiniGap} examined the scaling of the size of the mini-gap to sub-gap fermionic states localized, along with Majoranas, to the ends of perfectly rectangular $p+ip$ superconducting wires.  There it was found that the mini-gap, $\Delta_{\text{mg}}$ exhibited a maximum for wires with width $W\approx \xi_0$ which scaled as $\Delta_{\text{mg}}\approx \Delta_s^2/\e_F \ll \Delta_s$.  Qualitatively speaking, for perfectly rectangular wires, $\Delta_{\text{mg}}$ is very small because each sub-band contributes Majorana end-states which are nearly orthogonal to each other, and therefore mix only very weakly\cite{BrouewerMultiband}. 
\begin{figure}[hht]
\vspace{.1in}
\begin{center}
\hspace{-.3in}\includegraphics[width=3in]{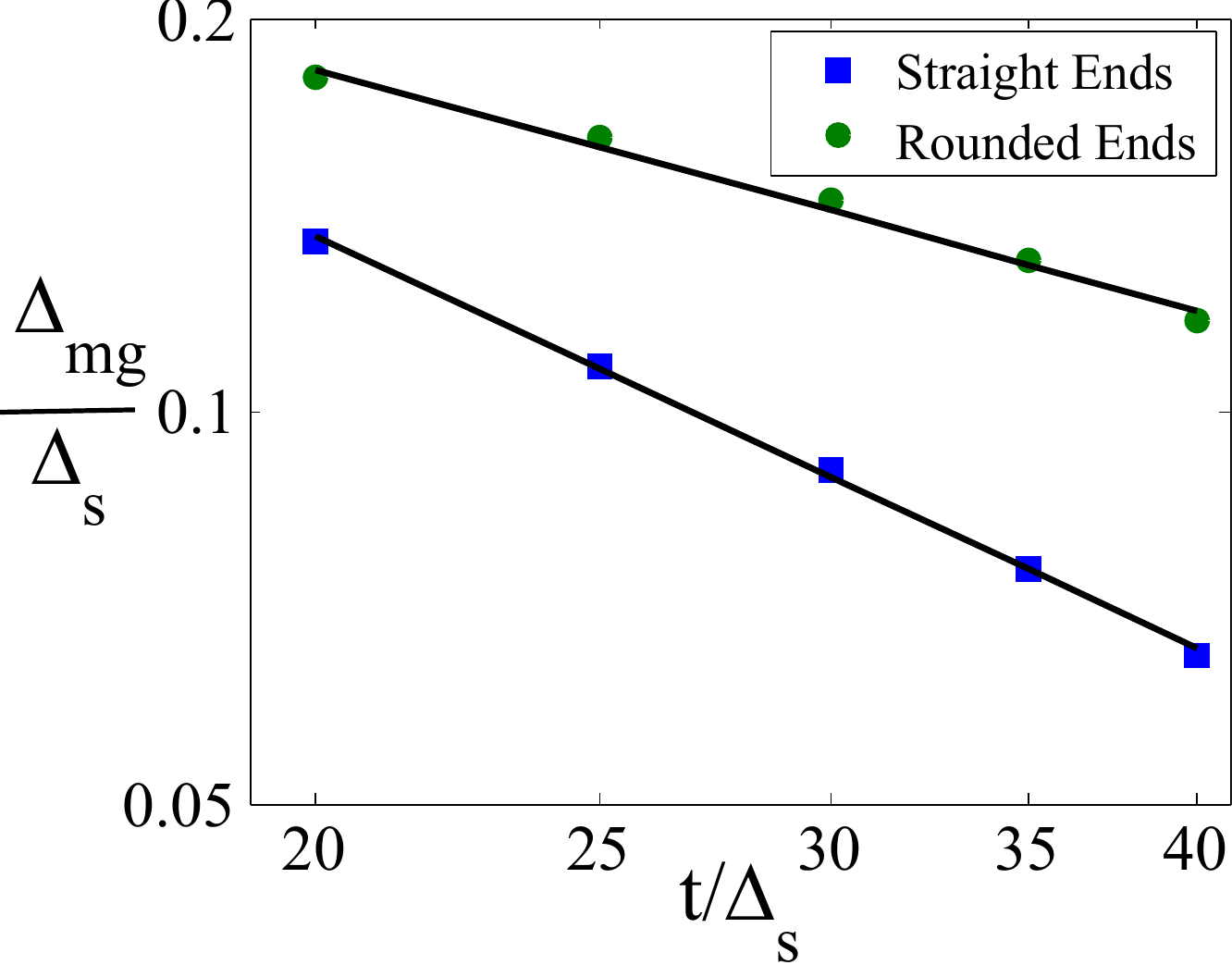}
\end{center}
\vspace{-.2in}
\caption{Log--log plot of scaling of maximal mini-gap, $\Delta_{\text{mg}}$, for the tight-binding model for $p+ip$ superconducting strips from Ref. \onlinecite{ACPMultiBand1} as a function of hopping strength, $t$.  The maximal mini-gap size occurs for strips of width $W\approx \xi_0$.  Best fit lines are shown in black.  For perfectly rectangular strips with straight ends (squares) $\Delta_{\text{mg}}\sim \Delta_s \(t/\Delta_s\)^{-1.052}$ as reported in \onlinecite{BrouwerMiniGap}.  In contrast wires with slightly rounded ends best--fit scales as $\Delta_{\text{mg}}\sim \Delta_s \(t/\Delta_s\)^{-0.614}$.}
\label{fig:MiniGapScaling}
\end{figure} 

However, this near perfect orthogonality is special to the case of perfectly rectangular sample geometry.  For perfect ends, the Majorana modes contributed by each sub-band are fine tuned to be almost exactly orthogonal to each--other.  In more realistic situations wire--ends (of either self-assembled semiconducting nanowires or microfabricated metallic strips) will not be so precise, giving rise to random overlaps between Majoranas from different sub-bands and leading.  In this appendix, we simulated strips with slightly distorted ends, and find a more favorable mini--gap scaling $\Delta_{\text{mg}}\approx \Delta_s\sqrt{\Delta_s/\e_F} \gg \Delta_s^2/\e_F$.  Specifically, we simulate numerically the tight-binding model for a $p+ip$ superconductor used in Refs. \onlinecite{ACPMultiBand1} and \onlinecite{BrouwerMiniGap}:
$H = H_{t} + H_{\text{p-BCS}}$, of a single species of electrons with $p_x+ip_y$ BCS pairing:
\begin{eqnarray} H_{t} &=& \sum_{\<ij\>} -t\(c^\dagger_ic_j + \text{h.c.}\)-\sum_j \mu c^\dagger_jc_j \nonumber\\
H_{\text{p-BCS}} &=& \sum_j \Delta_s\(-ic^\dagger_{j+\hat{x}}c^\dagger_j+c^\dagger_{j+\hat{y}}c^\dagger_j\)+\text{h.c} \label{eq:TBModel}\end{eqnarray}
where $c^\dagger_j$ creates an electron on site $j$,  $t$ is the hopping amplitude, $\mu$ is the chemical potential, $\Delta_s$ is the p-wave pairing amplitude, and we work in units where the lattice spacing is unity.
However, instead of rectangular strips, we consider nearly rectangular strips with elliptical capped ends.  When the length of the elliptical cap is larger than the Fermi wavelength, $1/k_F$, but still much smaller than the coherence length, $\xi_0$, the $\Delta_{\text{mg}}$ is parametrically enhanced.  

Fig. \ref{fig:MiniGapScaling} shows the optimal mini-gap scalings for wires with straight and rounded ends as a function of $\e_F\sim t$.  In these simulations, the length of the wire was chosen to be $L=10\xi_0$, the width was chosen as $W\approx \xi_0$ to optimize the $\Delta_{\text{mg}}$.  The length of the rounded elliptical cap was 5 lattice spacings, and the chemical potential was fixed at $\mu=-2t$.  The surface--pairing gap was chosen to be much less then $\e_F$ ($\Delta_s \ll t$), so that the coherence length was much longer than the lattice spacing. 

Qualitatively, we expect that the slightly rounded edges produce Majorana end-states for each sub-band which have the usual transverse profile along the width of the wire, are confined to the end of the wire with characteristic size $\xi_0$, and are randomly oscillating with wave-length $\approx k_F$ along the length of the wire.  The random oscillations along the wire give rise to random overlaps between different sub-bands, which based on the central limit theorem one would expect to scale as $\approx \sqrt{k_F\xi_0}$ in the limit $\frac{1/k_F}{\xi_0}\rightarrow \infty$.  The best-fit line in Fig. \ref{fig:MiniGapScaling} has a slightly different exponent ($\approx 0.6$ rather than the $0.5$ suggested by the above argument), which we expect is due to imperfect randomization by our choice of geometry as well as being limited to $\frac{1/k_F}{\xi_0}\sim 20-40$.


\begin{thebibliography}{99}
\bibitem{SubbandEndnote}
For $\mu<0$, the sub-band bottoms occur for $k_x\neq 0$ due to the ``mexican-hat" shape of the Rashba band.  Despite this, the topological phase transitions are still signaled by the gap closing at $k_x=0$.
\end{thebibliography}
\end{document}